\documentclass[%
 reprint,
superscriptaddress,
 amsmath,amssymb,
 aps,
pra,
]{revtex4-2}

\usepackage{graphicx}% Include figure files
\usepackage{dcolumn}% Align table columns on decimal point
\usepackage{bm}% bold math
\usepackage{gensymb}
\begin{document}

\preprint{APS/123-QED}

\title{A generalized approach to quantum interference in lossy $N$-port devices \\ via a singular value decomposition}

\author{Osmery Hernández}
\affiliation{%
 Department of Electrical, Electronic and Communications Engineering, Institute of Smart Cities (ISC), Public University of Navarre (UPNA), 31006 Pamplona, Spain
}%
\author{Iñigo Liberal}
\thanks{Corresponding author: inigo.liberal@unavarra.es}%
\affiliation{%
 Department of Electrical, Electronic and Communications Engineering, Institute of Smart Cities (ISC), Public University of Navarre (UPNA), 31006 Pamplona, Spain
}%

%\date{\today}% It is always \today, today,
             %  but any date may be explicitly specified

\begin{abstract}
Modeling quantum interference in the presence of dissipation is a critical aspect of quantum technologies. Including dissipation into the model of a linear device enables for assesing the detrimental impact of photon loss, as well as for studying dissipation-driven quantum state transformations. However, establishing the input-output relations characterizing quantum interference at a general lossy $N$-port network poses important theoretical challenges. Here, we propose a general procedure based on the singular value decomposition (SVD), which allows for the efficient calculation of the input-output relations for any arbitrary lossy linear device. In addition, we show how the SVD provides an intuitive description of the principle of operation of linear optical devices. We illustrate the applicability of our method by evaluating the input-output relations of popular reciprocal and nonreciprocal lossy linear devices, including devices with singular and nilpotent scattering matrices. We expect that our procedure will motivate future research on quantum interference in complex devices, as well as the realistic modelling of photon loss in linear lossy devices.
\end{abstract}

%\keywords{Suggested keywords}%Use showkeys class option if keyword
                              %display desired
\maketitle

%\tableofcontents

\section{Introduction}

Quantum interference - the superposition of nonclassical light states leading to the additive combination of their probability amplitudes - lies at the core of quantum technologies. In essence, combining nonclassical light states with a linear optical device enables quantum state transformation and entanglement generation, which are the basis for quantum metrology \cite{bouchard2020two, degen2017quantum,aguilar2020robust}, communication \cite{gisin2002quantum,sibson2017chip, sibson2017integrated}, simulation \cite{peruzzo2014variational,sparrow2018simulating}, and computing \cite{kok2007linear,zhong2020quantum} systems. Most research of quantum interference phenomena in linear optical systems has focused on lossless devices, with state-of-the-art architectures being composed of complicated networks of beamsplitters, Mach-Zehnder interferometers and directional couplers \cite{wang2020integrated,zhong2020quantum}. However, quantum interference phenomena in lossy devices might be equally interesting. First, all practical devices are lossy up some degree, and the detrimental impact of absorption must be critically assessed. This aspect is of particular relevance for quantum technologies, since their performance is more sensitive to photon loss than in classical systems. Second, absorption is an optical process of interest on its own right, with multiple applications in optics including sensing \cite{liu2010infrared,kravets2013singular}, thermophotovoltaics \cite{fleming2002all}, photodetection \cite{knight2011photodetection,krayer2019optoelectronic}, hot-electron chemistry \cite{mukherjee2013hot} and thermally-assisted optical tweezers \cite{ndukaife2016long}. Moreover, phase-dependent absorption, commonly known as coherent perfect absorption (CPA) \cite{Chong2010coherent, baranov2017coherent}, enables photon-photon interactions even in the absence of nonlinearities, of interest for lineal optical switching \cite{fang2014ultrafast,Zhao2016metawaveguide}, modulation \cite{xomalis2018fibre}, logical operations \cite{papaioannou2016two}, amplification \cite{fang2014ultrafast,fang2015controlling}, as well as quantum state transformations \cite{vetlugin2021coherent}.

However, modeling quantum systems in the presence of dissipation poses important theoretical challenges. Typically, the quantum mechanical description of the electromagnetic field must be complemented by coupling it to a continuum of polaritonic modes, representing the excitations in the matter system, as well as irreversible dissipation \cite{vogel2006quantum,scheel2009macroscopic,philbin2010canonical, drummond1990electromagnetic,huttner1992quantization}. The coupling between the field and matter systems must be correctly defined in order to ensure that the macroscopic response of a lossy material is recovered. A simplified procedure can be carried out for linear optical devices with well-defined input and output ports. In such a case, a reduced set of polaritonic modes can describe the internal degrees of freedom of the device, as well as the dissipation within it \cite{barnett1998quantum, jeffers2000interference, gruner1996quantum, knoll1999quantum}. Following these models, the response of a lossy device is described in terms of generalized input-output relations, directly linked to its classical transmission and reflection coefficients. Such “black box” approach facilitates a compact and efficient evaluation of quantum interference phenomena. In fact, these models successfully describe a variety of recent experiments, such as the anti-coalescence of photons and nonlinear absorption \cite{vest2017anti, vetlugin2021anti}, single-photon coherent perfect absorption \cite{roger2015coherent, vest2018plasmonic, vetlugin2019coherent} and coherent absorption of N00N states \cite{roger2016coherent, lyons2019coherent}. In addition, input-output theory of lossy beam splitters has inspired interesting theoretical proposals on nonlocal absorption \cite{jeffers2019nonlocal} and quantum coherent absorption of squeezed light \cite{hardal2019quantum}.

Previous works on quantum interference in lossy linear devices have been mostly restricted to the analysis of lossy beam splitters. However, it should be expected that investigating more complex devices will lead to the discovery of new forms of quantum interference. At the same time, the calculation of input-output relations for arbitrary lossy $N$-port devices will require from advanced modeling techniques. Here, we present a systematic procedure that allows for the calculation of the input-output relations for any arbitrary lossy $N$-port device, thus allowing for a full characterization of quantum interference phenomena in the presence of loss. Our procedure is based on using a singular value decomposition (SVD) \cite{gentle2012numerical, golub2013matrix, horn2012matrix, datta2004numerical, lindfield2018numerical} for the matrices that describe the transformation of optical modes, as well as the dissipation into the device. As we will show, the SVD is a powerful tool that provides a very intuitive picture of how the device operates. Moreover, our procedure can be applied to any lossy $N$-port device, e.g., reciprocal and nonreciprocal devices, and devices with singular scattering matrices. We illustrate the applicability of our method by deriving the input-output relations for several reciprocal and nonreciprocal devices such as lossy T-junctions, Wilkinson power dividers/combiners, circulators and isolators. We expect that our results will motivate future research of quantum interference phenomena in complex $N$-port devices.

\section{Quantum interference in lossy $N$-port networks}

\begin{figure}[!t]
\includegraphics[width=3.35in]{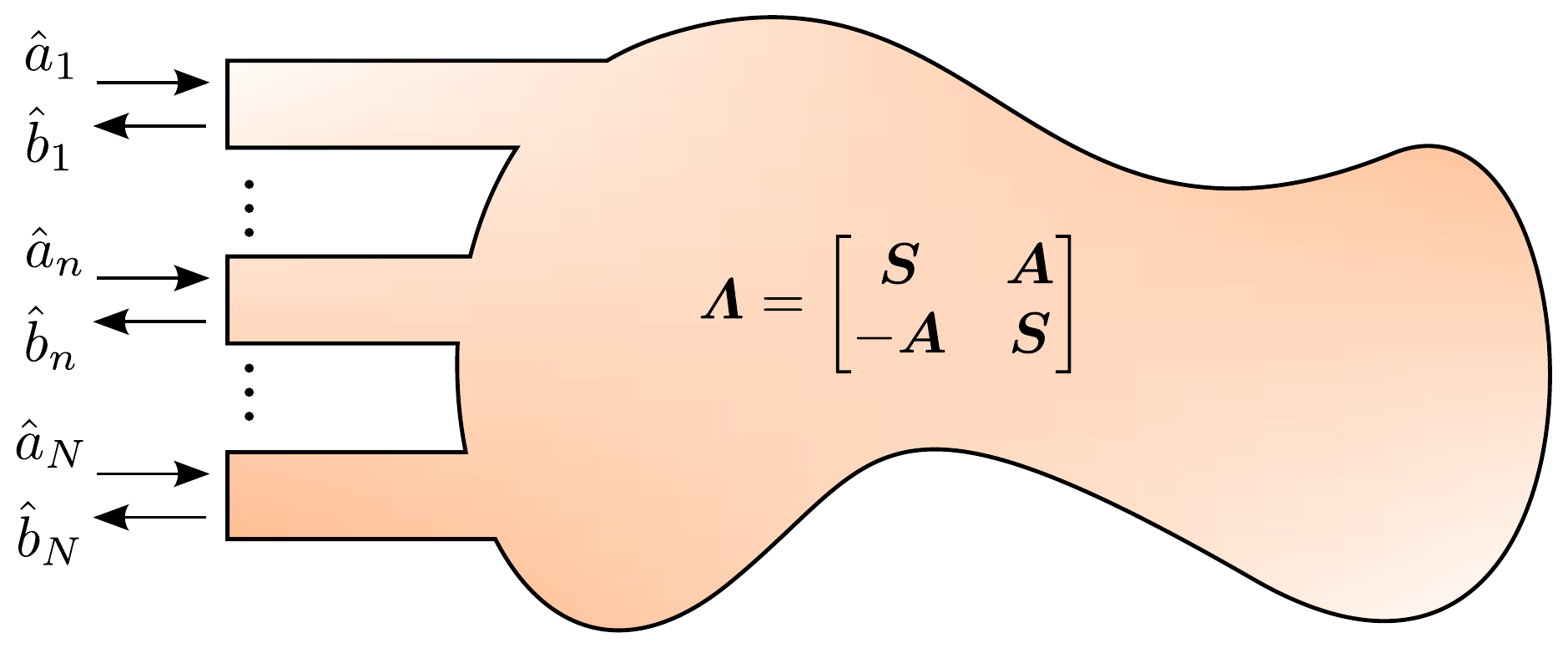}
\caption{Sketch of input-output relations in a general lossy $N$-port linear device.}
\label{fig:Main_result}
\end{figure}

Classical interference phenomena at an $N$-port device can be described through its scattering matrix, $\mathbf{S}\in\mathbb{C}^{N\times N}$, which provides an algebraic relation between the input and output waves: $\mathbf{b}=\mathbf{S}\,\mathbf{a}$, where $\mathbf{a}=\left[a_{1},\ldots,a_{N}\right]^{T}$ is a vector containing complex numbers describing the input waves, while $\mathbf{b}=\left[b_{1},\ldots,b_{N}\right]^{T}$ is the vector of the output waves. If the device is lossless, its scattering matrix is unitary, $\mathbf{S}\mathbf{S}^{\dagger}=\mathbf{I}$. For a lossless device, quantum interference phenomena is described by the same matrix. In this case, the scattering matrix provides input-output relations for the photonic destruction operators (see Fig.~\ref{fig:Main_result}) 
\begin{equation}
\widehat{\mathbf{b}}\,=\,\mathbf{S}\,\,\widehat{\mathbf{a}}
\label{eq:IO_lossless}
\end{equation}

\noindent where $\widehat{\mathbf{a}}=\left[\widehat{a}_{1},\ldots,\widehat{a}_{N}\right]^{T}$ and $\widehat{\mathbf{b}}=\left[\widehat{b}_{1},\ldots,\widehat{b}_{N}\right]^{T}$ are now vectors of input and output operators obeying bosonic commutation relations: $\left[\hat{a}_{n},\hat{a}_{m}^{\dagger}\right]=\left[\hat{b}_{n},\hat{b}_{m}^{\dagger}\right]=\delta_{nm}$, and $\left[\hat{a}_{n},\hat{a}_{m}\right]=\left[\hat{b}_{n},\hat{b}_{m}\right]=0$. 

%%%%%%%%%%%%%%%%%%%%%%%%%%%%%%%%%%%%%%%%% Figure 1 %%%%%%%%%%%%%%%%%%%%%%%%%%%%%%%%%%%%%%%%%%%%%%%%%%%%%%

%%%%%%%%%%%%%%%%%%%%%%%%%%%%%%%%%%%%%%%%%%%%%%%%%%%%%%%%%%%%%%%%%%%%%%%%%%%%%%%%%%%%%%%%%%%%%%%%%%%%%%%%%

The description of quantum interference in a lossy device is a more complicated task. As anticipated, the input-output relations need to be generalized \cite{barnett1998quantum, jeffers2000interference, gruner1996quantum, knoll1999quantum}. To this end, we define input $\hat{\boldsymbol{f}}=\left[\hat{f}_{1},...,\hat{f}_{N}\right]^{T}$ and output $\hat{\boldsymbol{g}}=\left[\hat{g}_{1},...,\hat{g}_{N}\right]^{T}$ bosonic operators describing the internal degrees of freedom of the device. In this manner, the scattering matrix $\boldsymbol{S}$ is expanded into a more general matrix, $\boldsymbol{\varLambda}$, providing generalized input-output relations for a lossy device:
\begin{equation}
\begin{bmatrix}\hat{\boldsymbol{b}}\\
\hat{\boldsymbol{g}}
\end{bmatrix}=\boldsymbol{\varLambda}\begin{bmatrix}\hat{\boldsymbol{a}}\\
\hat{\boldsymbol{f}}
\end{bmatrix}
\label{eq:input-output}
\end{equation}

The $\boldsymbol{\varLambda}$ matrix fully describe the input-output relations. In doing so, it contains all the information required to evaluate quantum interference phenomena. In order to determine the generalized matrix $\boldsymbol{\varLambda}$, it is convenient to write it as a block matrix of the following form \cite{knoll1999quantum}:
\begin{equation}
\boldsymbol{\varLambda}=\begin{bmatrix}\boldsymbol{S} & \boldsymbol{A}\\
\boldsymbol{C} & \boldsymbol{D}
\end{bmatrix}
\label{eq:Delta}
\end{equation}

\noindent where $\boldsymbol{S}$ and $\boldsymbol{A}$ are respectively the scattering and the absorption matrices, while $\boldsymbol{C}$ and $\boldsymbol{D}$ can be written in terms of $\boldsymbol{S}$ and $\boldsymbol{A}$, as demonstrated below. $\boldsymbol{\varLambda}$ must be a unitary matrix ($\boldsymbol{\varLambda}\boldsymbol{\varLambda}^{\dagger}=\boldsymbol{I}$) since a linear device preserves the number of excitations. Therefore, the matrices $\boldsymbol{S}$, $\boldsymbol{A}$, $\boldsymbol{C}$ and $\boldsymbol{D}$ must satisfy the following relations:
\begin{equation}
\boldsymbol{S}\boldsymbol{S}^{\dagger}+\boldsymbol{A}\boldsymbol{A}^{\dagger}=\boldsymbol{I}
\label{eq:SS_AA_I}
\end{equation}
\begin{equation}
\boldsymbol{C}\boldsymbol{C}^{\dagger}+\boldsymbol{D}\boldsymbol{D}^{\dagger}=\boldsymbol{I}
\label{eq:CC_DD_I}
\end{equation}
\begin{equation}
\boldsymbol{S}\boldsymbol{C}^{\dagger}+\boldsymbol{A}\boldsymbol{D}^{\dagger}=\boldsymbol{0}
\label{eq:SC_AD_0}
\end{equation}

The same conditions can be independently derived by imposing the bosonic commutation relations for the output operators:
\begin{equation}
\left[\hat{b}_{n},\hat{b}_{m}^{\dagger}\right]=\mathop{\sum_{p=1}^{N}}S_{np}S_{mp}^{*}+A_{np}A_{mp}^{*}=\delta_{nm}
\label{eq:Comm_OutputOper}
\end{equation}
\begin{equation}
\left[\hat{g}_{n},\hat{g}_{m}^{\dagger}\right]=\mathop{\sum_{p=1}^{N}}C_{np}C_{mp}^{*}+D_{np}D_{mp}^{*}=\delta_{nm}
\label{eq:Comm_NoiseOper}
\end{equation}
\begin{equation}
\left[\hat{b}_{n},\hat{g}_{n}^{\dagger}\right]=\mathop{\sum_{p=1}^{N}}S_{np}C_{np}^{*}+A_{np}D_{np}^{*}=0
\label{eq:Comm_OutputNoiseOper}
\end{equation}

It can be readily checked that the constraints given by Eqs.~(\ref{eq:Comm_OutputOper})-(\ref{eq:Comm_OutputNoiseOper}) are fully equivalent to those in Eqs.~(\ref{eq:SS_AA_I})-(\ref{eq:SC_AD_0}). Therefore, we can understand the requirement of the unitary property of matrix $\boldsymbol{\varLambda}$ as being the result of both: (i) the conservation of the number of quantum excitations or (ii) the conservation of the bosonic nature of the quantum operators.

Since $\boldsymbol{\varLambda}$ is unitary, we have $\boldsymbol{\varLambda}^{-1}=\boldsymbol{\varLambda}^{\dagger}$ leading to the inverse replacement rules
\begin{equation}
\widehat{a}_{n}^{\dagger}\rightarrow\sum_{m=1}^{M}\,\left(S_{mn}\widehat{b}_{m}^{\dagger}+C_{mn}\widehat{g}_{m}^{\dagger}\right)
\label{eq:a_dagger}
\end{equation}
\begin{equation}
\widehat{f}_{n}^{\dagger}\rightarrow\sum_{m=1}^{M}\,\left(A_{mn}\widehat{b}_{m}^{\dagger}+D_{mn}\widehat{g}_{m}^{\dagger}\right)
\label{eq:f_dagger}
\end{equation}

Using (\ref{eq:a_dagger})-(\ref{eq:f_dagger}), the state at the output of a linear optical device can be readily determined. In particular, the input state can be written as general function of creation operators with the input operators: 
\begin{equation}
\left|\psi_{\mathrm{in}}\right\rangle =F\left(\widehat{a}_{1}^{\dagger},\ldots,\widehat{a}_{N}^{\dagger};\widehat{f}_{1}^{\dagger},\ldots,\widehat{f}_{N}^{\dagger}\right)\left|0\right\rangle 
\label{eq:psi_in}
\end{equation}

\noindent and the output state $\left|\psi_{\mathrm{out}}\right\rangle$  is determined by substituting (\ref{eq:a_dagger})-(\ref{eq:f_dagger}) into (\ref{eq:psi_in}).

A typical scenario in studying quantum interference in lossy devices is starting with a device with a known scattering matrix, $\mathbf{S}$, which can be determined from the classical characterization of the device and/or numerical simulations. Then, one has to identify the $\mathbf{A}$,\,$\mathbf{C}$ and $\mathbf{D}$ matrices that compose the complete $\boldsymbol{\varLambda}$ matrix, fully characterizing quantum interference. In \cite{knoll1999quantum} a general solution for calculating the matrix $\boldsymbol{\varLambda}$ of lossy two-port devices has been proposed. The method can be extrapolated to $N$-port devices. However, the solution requires from matrix inversion at several steps, and that it cannot be applied to devices with singular matrices such as isolators. Recent works have shown that the $\boldsymbol{\varLambda}$ matrix reduces to a simpler form for devices with a real and symmetric scattering matrix \cite{hardal2019quantum}. In the following sections, we proposed the singular value decomposition (SVD) as a general procedure that enables the description of any arbitrary linear device in a very compact form.

\section{Computation of input-output relations with a singular value decomposition (SVD)}

The main contribution of this work is that the $\boldsymbol{\varLambda}$ matrix for an $N$-port device can be written in the following simple and compact form, for all classes of linear devices:
\begin{equation}
\boldsymbol{\varLambda}=\begin{bmatrix}\boldsymbol{S} & \boldsymbol{A}\\
-\boldsymbol{A} & \boldsymbol{S}
\end{bmatrix}
\label{eq:Delta_SVD}
\end{equation}

Specifically, Eq.~(\ref{eq:Delta_SVD}) can be derived by applying a singular value decomposition (SVD) to the scattering matrix $\boldsymbol{S}$, and selecting matrices $\boldsymbol{A}$,$\mathbf{C}$ and $\mathbf{D}$ that share the same factorization. To demonstrate that this is the case, we start by writing the $\boldsymbol{S}$ matrix on its SVD form. The singular value decomposition (SVD) \cite{gentle2012numerical, golub2013matrix, horn2012matrix, datta2004numerical, lindfield2018numerical} is a powerful method that allows for the factorization of any rectangular matrix. However, since we focus on matrices characterizing $N$-port devices, our derivation only needs to apply to square matrices, $\boldsymbol{S}\in\mathbb{C}^{N\times N}$. The SVD states that any general complex matrix $\boldsymbol{S}$ can be factorized as follows
\begin{equation}
\boldsymbol{S}=\boldsymbol{U}\boldsymbol{\Sigma}_{S}\boldsymbol{V}^{\dagger}
\label{eq:S_SVD}
\end{equation}

All three matrices involved in the factorization are square matrices, $\boldsymbol{U}$,$\boldsymbol{\Sigma}_{S}$,$\boldsymbol{V}\in\mathbb{C}^{N\times N}$. First, $\boldsymbol{\Sigma}_{S}=diag\left\{ d_{1},...,d_{N}\right\}$  is a diagonal matrix whose entries are the singular values of $\boldsymbol{S}$, i.e., real numbers corresponding to the nonnegative square roots of the eigenvalues of $\boldsymbol{S}\boldsymbol{S}^{\dagger}$, arranged in decreasing order. Its values are also equal or smaller than one, since the device is passive, i.e., $d_{n}\in\left[0\,,\,1\right]\:\forall\,n$. $\boldsymbol{U}$ is a unitary matrix $\boldsymbol{U}\boldsymbol{U}^{\dagger}=\mathbf{I}$ whose columns are the eigenvectors of the matrix $\boldsymbol{S}\boldsymbol{S}^{\dagger}$, also known as left singular vectors. Similarly, $\boldsymbol{V}$ is a unitary matrix $\boldsymbol{V}\boldsymbol{V}^{\dagger}=\mathbf{I}$ whose columns are the eigenvectors of the matrix $\boldsymbol{S}^{\dagger}\boldsymbol{S}$, also called right singular vectors \cite{gentle2012numerical, golub2013matrix, horn2012matrix, datta2004numerical, lindfield2018numerical}. Note that the SVD is not strictly unique. Although the matrix $\boldsymbol{\Sigma}_{S}$ is uniquely defined by the unique singular values of the $\boldsymbol{S}$ matrix, the matrices $\boldsymbol{U}$ and $\boldsymbol{V}$ are only unique up to a phase factor \cite{datta2004numerical}.

The SVD provides a useful tool to intuitively understand the response of a device. The SVD form of the scattering matrix can be visualized as a one-to-one mapping from elements of one basis of $\mathbb{C}^{N}$, given by the columns of $\boldsymbol{V}$, to elements on a different basis of $\mathbb{C}^{N}$, given by the columns of $\boldsymbol{U}$, with a scalar mapping factor that is real and nonnegative, given by the diagonal entries of $\boldsymbol{\Sigma}_{S}$. In this manner, the response to a given vectorial input can be visualized based on how it projects onto the basis expanded by $\boldsymbol{V}$ and its transformation into the basis projected by $\boldsymbol{U}$. This procedure is similar to an eigendecomposition, except that the eigendecomposition maps eigenvectors onto a scaled version of themselves, with the scale factor being an in general complex eigenvalue. We will show that the SVD provides a very intuitive perspective on the mode of operation of linear optical devices. Many optical devices operate as transmission devices, where the input from one port propagates through the device and appears transmitted into the ports, while minimizing the reflection at the input port. With the reflection at the input port being zero, this mode of operation cannot be described via eigenvectors. Therefore, the use of two different basis, as it is the case in the SVD, is particularly useful to describe the mode of operation of transmission devices. Moreover, the most important advantage of the SVD is that it can be applied to any matrix. Therefore, the method introduced here can be applied to any linear device, for instance, reciprocal and nonreciprocal devices, extreme devices with real and complex matrices, devices with singular and nilpotent scattering matrices, etc.

Next, we use the SVD form of the scattering matrix $\boldsymbol{S}$ to solve for the conditions given by Eqs.\,(\ref{eq:SS_AA_I})-(\ref{eq:SC_AD_0}). First, we identify the absorption matrix $\mathbf{A}$ from condition (\ref{eq:SS_AA_I}), which can be rewritten as 
\begin{equation}
\boldsymbol{A}\boldsymbol{A}^{\dagger}=\boldsymbol{I}-\boldsymbol{S}\boldsymbol{S}^{\dagger}=\boldsymbol{I}-\left(\boldsymbol{U}\boldsymbol{\Sigma}_{S}\boldsymbol{V}^{\dagger}\right)\left(\boldsymbol{V}\boldsymbol{\Sigma}_{S}\boldsymbol{U}^{\dagger}\right)\nonumber\\
\end{equation}
\begin{equation}
=\boldsymbol{I}-\boldsymbol{U}\boldsymbol{\Sigma}_{S}^{2}\boldsymbol{U}^{\dagger}=\boldsymbol{U}\left(\boldsymbol{I}-\boldsymbol{\Sigma}_{S}^{2}\right)\boldsymbol{U}^{\dagger}\nonumber\\
\end{equation}
\begin{equation}
=\boldsymbol{U}\boldsymbol{\Sigma}_{A}\left(\boldsymbol{U}\boldsymbol{\Sigma}_{A}\right)^{\dagger}
\label{eq:AA}
\end{equation}

\noindent where we have defined the diagonal matrix $\boldsymbol{\Sigma}_{A}=diag\left\{ \sqrt{1-d_{1}^{2}},...,\sqrt{1-d_{N}^{2}}\right\}$. It is clear from Eq.~(\ref{eq:AA}) that the matrix $\boldsymbol{A}$ can be identified as $\boldsymbol{A}=\boldsymbol{U}\boldsymbol{\Sigma}_{A}$. However, it must be noted that the matrix $\boldsymbol{A}$ is not uniquely defined by condition (\ref{eq:SS_AA_I}), since any unitary transformation $\boldsymbol{A}'=\boldsymbol{A}\boldsymbol{U}_{A}^{\dagger}$ does not change the product $\boldsymbol{A}\boldsymbol{A}^{\dagger}$. The freedom in choosing the $\boldsymbol{A}$ matrix physically represents the possibility of changing the basis on which the internal modes of the device are represented. A natural choice might seem to choose $\boldsymbol{U}_{A}=\boldsymbol{U}$, so that $\boldsymbol{A}$ reduces to a Hermitian, positive semidefinite matrix. However, we suggest the use of $\boldsymbol{U}_{A}=\mathbf{V}$, so that $\boldsymbol{A}$ and $\boldsymbol{S}$ are factorized by the same unitary matrices:
\begin{equation}
\boldsymbol{A}=\boldsymbol{U}\boldsymbol{\Sigma}_{A}\boldsymbol{V}^{\dagger}
\label{eq:A_svd}
\end{equation}

Finally, we must identify matrices $\boldsymbol{C}$ and $\boldsymbol{D}$ that fulfill the conditions (\ref{eq:CC_DD_I}) and (\ref{eq:SC_AD_0}). Again, we opt for matrices that are factorized by the same unitary matrices: $\boldsymbol{C}=\boldsymbol{U}\boldsymbol{\Sigma}_{C}\boldsymbol{V}^{\dagger}$ and $\boldsymbol{D}=\boldsymbol{U}\boldsymbol{\Sigma}_{D}\boldsymbol{V}^{\dagger}$. Introducing these forms into (\ref{eq:CC_DD_I}) and (\ref{eq:SC_AD_0}), we can readily derive the conditions for the diagonal matrices $\boldsymbol{\Sigma}_{C}=-\boldsymbol{\Sigma}_{A}$ and $\boldsymbol{\Sigma}_{D}=\boldsymbol{\Sigma}_{S}$, leading to
\begin{equation}
\boldsymbol{C}=-\boldsymbol{A}
\label{eq:C_scd}
\end{equation}
\begin{equation}
\boldsymbol{D}=\boldsymbol{S}
\label{eq:D_svd}
\end{equation}

Combining (\ref{eq:A_svd}), (\ref{eq:C_scd}) and (\ref{eq:D_svd}) we arrive at the compact form for the $\boldsymbol{\varLambda}$ matrix for a lossy $N$-port network anticipated by Eq.~(\ref{eq:Delta_SVD}). It is straightforward to verify the unitary property of $\boldsymbol{\varLambda}$ through Eq.~(\ref{eq:Delta_SVD}) and the decomposition of $\boldsymbol{S}$ and $\boldsymbol{A}$ via Eqs.~(\ref{eq:S_SVD}) and (\ref{eq:A_svd}), respectively.

We re-emphasize that a major advantage of the described procedure is that it can be applied to any lossy device since any general complex scattering matrix $\boldsymbol{S}$, regardless of its type, admits a singular value decomposition. Moreover, popular mathematical software packages provide simple commands and routines to carry out the SVD \cite{Matlab, Mathematica}, so that it can be easily computed and/or incorporated into design routines. Finally, we note that although the SVD can be applied to any matrix, it reduces to a simpler form in some specific cases. For example, if the scattering matrix is real, $\boldsymbol{S}\in\mathbb{\mathbb{R}}^{N\times N}$, then the three matrices involved in the decomposition in Eq.~(\ref{eq:S_SVD}) will have only real elements. Hence, $\boldsymbol{U}$ and $\boldsymbol{V}$ will be orthogonal matrices and the Hermitian adjoint operation acting on $\boldsymbol{V}$ will be reduced to a transposition operation, $\boldsymbol{S}=\boldsymbol{U}\boldsymbol{\Sigma}_{S}\mathbf{V}^{T}$. On the other hand, if the scattering matrix is a complex symmetric matrix $\boldsymbol{S}=\boldsymbol{S}^{T}$, then there is a choice for the unitary matrices such that $\boldsymbol{V}=\boldsymbol{U}^{*}$. Therefore, $\boldsymbol{S}=\boldsymbol{U}\boldsymbol{\Sigma}_{S}\boldsymbol{U}^{T}$, a result known as the Autonne-Takagi factorization \cite{horn2012matrix}. In practice, the scattering matrix of a reciprocal device is symmetric, so all reciprocal devices can be described through the Autonne-Takagi factorization.

\section{Computation of input-output relations with a unitary diagonalization}

Although the SVD provides a general technique that can be applied to any linear device, other approaches might also be considered in particular cases. For example, the characterization of the response of a system by identifying its eigenvectors is ubiquitous in all branches of science and technology. Therefore, a factorization based on eigenvectors might provide additional insight in many circumstances. To address this point, we consider the diagonalization by a unitary similarity transformation, in which the scattering matrix $\boldsymbol{S}$ is factorized as follows
\begin{equation}
\boldsymbol{S}=\boldsymbol{U}_{\lambda}\boldsymbol{\Sigma}_{\lambda}\boldsymbol{U}_{\lambda}^{\dagger}
\label{eq:S_UDiagon}
\end{equation}

\noindent where $\boldsymbol{\Sigma}_{\lambda}=diag\left\{ \lambda_{1},...,\lambda_{N}\right\}$ is a diagonal matrix whose entries are the eigenvalues of $\boldsymbol{S}$, while $\boldsymbol{U}_{\lambda}$ is a unitary matrix whose columns are eigenvectors of $\boldsymbol{S}$. Eigenvectors map onto a scaled version of themselves through the scattering matrix, which can alternatively provide an intuitive picture of how a device reacts to external excitations. Unfortunately, the unitary diagonalization applies exclusively to normal matrices, and it therefore has a more limited scope. By definition, normal matrices are those which commute with their Hermitian adjoint, i.e., $\left[\boldsymbol{S},\boldsymbol{S}^{\dagger}\right]=\boldsymbol{S}\boldsymbol{S}^{\dagger}-\boldsymbol{S}^{\dagger}\boldsymbol{S}=0$. Hermitian, anti-Hermitian, unitary, orthogonal and symmetric matrices are typical examples of normal matrices, which can be found in a large number of physical scenarios \cite{macklin1984normal}. If the scattering matrix of a given device is identified to belong to one of those classes of matrices, then an analysis via diagonalization by unitary transformations can be performed.

The use of this decomposition leads to some minor changes in the proposed method to calculate matrix $\boldsymbol{\varLambda}$, since the diagonal entries in $\boldsymbol{\Sigma}_{\lambda}$ are not necessary real nonnegative values. Aside from that, one can proceed with the derivation in a very similar manner. First, we assume that the $\boldsymbol{A}_{\lambda}=\boldsymbol{U}_{\lambda}\boldsymbol{\Sigma}_{A\lambda}\boldsymbol{U}_{\lambda}^{\dagger}$, $\boldsymbol{C}_{\lambda}=\boldsymbol{U}_{\lambda}\boldsymbol{\Sigma}_{C\lambda}\boldsymbol{U}_{\lambda}^{\dagger}$ and $\boldsymbol{D}_{\lambda}=\boldsymbol{U}_{\lambda}\boldsymbol{\Sigma}_{D\lambda}\boldsymbol{U}_{\lambda}^{\dagger}$ matrices are diagonalizable by the same unitary matrix. Then, the diagonal matrix solving conditions (\ref{eq:SS_AA_I})-(\ref{eq:SC_AD_0}) are found to be $\boldsymbol{\Sigma}_{A\lambda}=diag\left\{ \sqrt{1-\left|\lambda_{1}\right|^{2}},...,\sqrt{1-\left|\lambda_{N}\right|^{2}}\right\}$, $\boldsymbol{\Sigma}_{C\lambda}=-\boldsymbol{\Sigma}_{A\lambda}$ and $\boldsymbol{\Sigma}_{D\lambda}=\boldsymbol{\Sigma}_{\lambda}^{*}$. In this manner, the matrix $\boldsymbol{\varLambda}$ takes the following form:
\begin{equation}
\boldsymbol{\varLambda}=\begin{bmatrix}\boldsymbol{S} & \boldsymbol{A}_{\lambda}\\
-\boldsymbol{A}_{\lambda} & \boldsymbol{S}^{\dagger}
\end{bmatrix}
\label{eq:Delta_UDiagon}
\end{equation}

To finalize, it is worth remarking that there is a direct and closed relationship between the singular value decomposition and the unitary diagonalization. Since the singular values correspond to the absolute values of the eigenvalues ($d_{n}=\left|\lambda_{n}\right|$), it is straightforward to obtain an SVD decomposition of a normal matrix from its unitary diagonalization. To this end, if the $n^{th}$-eigenvalue is defined as $\lambda_{n}=\left|\lambda_{n}\right|e^{i\vartheta_{n}}$ we can write $\boldsymbol{\Sigma}_{\lambda}=\boldsymbol{\Sigma}_{\vartheta}\boldsymbol{\Sigma}_{\left|\lambda\right|}$, where $\boldsymbol{\Sigma}_{\vartheta}=diag\left\{ e^{i\vartheta_{1}},...,e^{i\vartheta_{N}}\right\}$  and $\boldsymbol{\Sigma}_{\left|\lambda\right|}=diag\left\{ \left|\lambda_{1}\right|,...,\left|\lambda_{N}\right|\right\}$. Then, Eq.~(\ref{eq:S_UDiagon}) can be rearranged as $\boldsymbol{S}=\left(\boldsymbol{U}_{\lambda}\boldsymbol{\Sigma}_{\vartheta}\right)\boldsymbol{\Sigma}_{\left|\lambda\right|}\boldsymbol{U}_{\lambda}^{\dagger}$ which is an SVD of $\boldsymbol{S}$ with $\boldsymbol{U}=\boldsymbol{U}_{\lambda}\boldsymbol{\Sigma}_{\vartheta}$, $\boldsymbol{\Sigma}_{S}=\boldsymbol{\Sigma}_{\left|\lambda\right|}$ and $\boldsymbol{V}=\boldsymbol{U}_{\lambda}$. For the same reason, it follows that $\boldsymbol{\Sigma}_{A\lambda}=\boldsymbol{\Sigma}_{A}$.

\section{Examples}

Once we have outlined a general method to compute the $\boldsymbol{\varLambda}$ matrix, we demonstrate its applicability calculating the input-output relations for several examples of popular linear devices, both reciprocal and nonreciprocal. In these examples, the SVD and unitary similarity diagonalizations are computed by means of a commercial mathematical package \cite{Mathematica}.

\subsection{Reciprocal devices}

\subsubsection{Lossy T-junction power divider}

\begin{figure}[!t]
\includegraphics[width=3.35in]{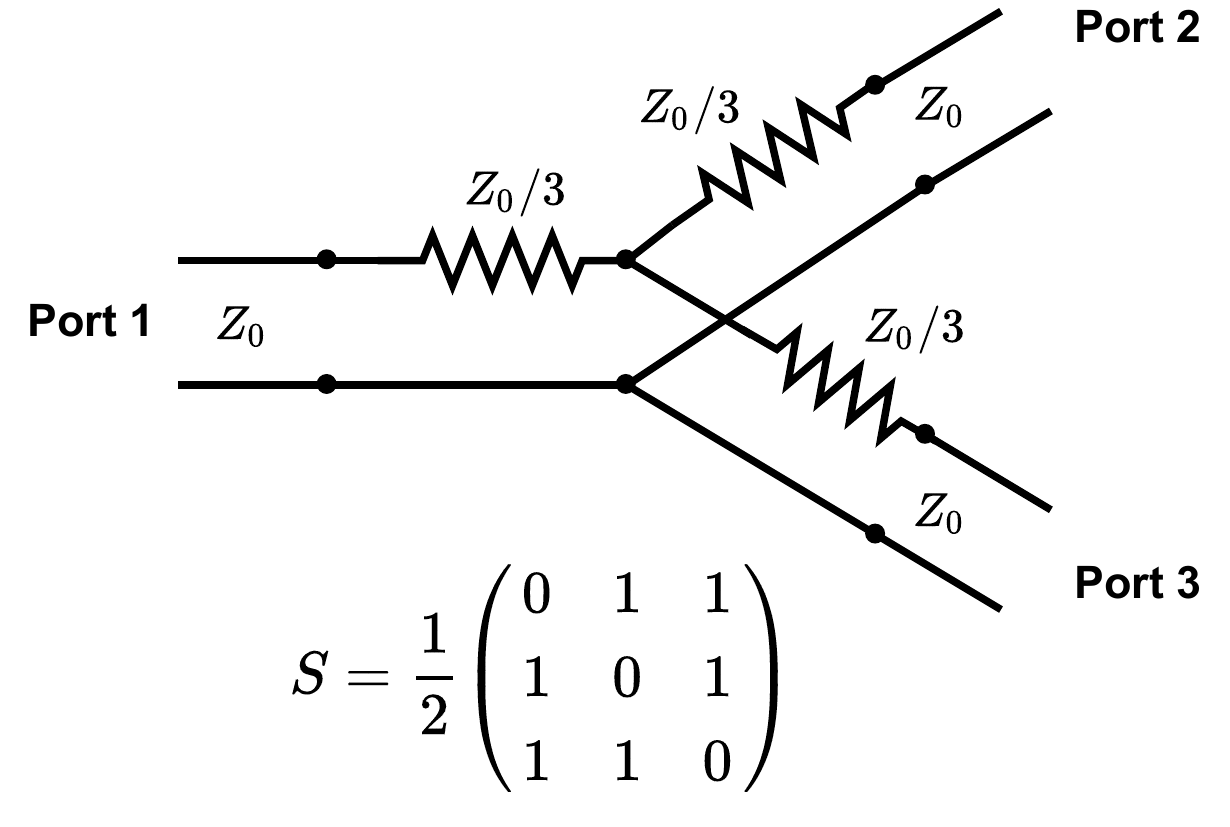}
\caption{Transmission line circuit and scattering matrix of a lossy T-junction power divider with an equal power splitting ratio.}
\label{fig:Lossy_TJunction_divider}
\end{figure}

As a first example of the application of the proposed method to compute input-output relations, we study the case of a lossy T-junction power divider, whose schematic representation and scattering matrix are depicted in Fig.~\ref{fig:Lossy_TJunction_divider}. The functionalities of a lossy T-junction power divider can be inferred from its scattering matrix. It consists of a three-port device with all its ports matched ($S_{ii}=0$), that performs power division and power combining with an equal power splitting ratio ($S_{ij}=1/2$, $\forall i\neq j$). Furthermore, since it is a reciprocal device, its $\boldsymbol{S}$ matrix is symmetric ($S_{ij}=S_{ji}$). Therefore, lossy T-junction power dividers provide the functionalities of power splitting and power combining, while guaranteeing that there are no back reflections. Moreover, all ports have exactly the same response, so that the device can be operated in any direction. However, these desired characteristics are achieved at the cost of residual absorption and a low transmission efficiency \cite{pozar2011microwave}.

%%%%%%%%%%%%%%%%%%%%%%%%%%%%%%%%% Figure 2 %%%%%%%%%%%%%%%%%%%%%%%%%%%%%%%%%%%%%%%%%%%%%%%%%%%%%%%%%%%%%

%%%%%%%%%%%%%%%%%%%%%%%%%%%%%%%%%%%%%%%%%%%%%%%%%%%%%%%%%%%%%%%%%%%%%%%%%%%%%%%%%%%%%%%%%%%%%%%%%%%%%%%%

Next, we obtain the $\boldsymbol{\varLambda}$ matrix via a singular value decomposition of $\boldsymbol{S}$, i.e., $\boldsymbol{S}=\boldsymbol{U}\boldsymbol{\Sigma}_{S}\boldsymbol{V}^{\dagger}$. For the scattering matrix reported in Fig.~\ref{fig:Lossy_TJunction_divider}, the involved matrices are given by
\begin{equation}
\boldsymbol{U}=\begin{bmatrix}\frac{1}{\sqrt{3}} & \frac{1}{\sqrt{2}} & \frac{1}{\sqrt{6}}\\
\frac{1}{\sqrt{3}} & 0 & -\sqrt{\frac{2}{3}}\\
\frac{1}{\sqrt{3}} & -\frac{1}{\sqrt{2}} & \frac{1}{\sqrt{6}}
\end{bmatrix}\qquad\boldsymbol{\Sigma}_{S}=\begin{bmatrix}1 & 0 & 0\\
0 & \frac{1}{2} & 0\\
0 & 0 & \frac{1}{2}
\end{bmatrix}\nonumber
\end{equation}
\begin{equation}
\boldsymbol{V}=\begin{bmatrix}\frac{1}{\sqrt{3}} & -\frac{1}{\sqrt{2}} & -\frac{1}{\sqrt{6}}\\
\frac{1}{\sqrt{3}} & 0 & \sqrt{\frac{2}{3}}\\
\frac{1}{\sqrt{3}} & \frac{1}{\sqrt{2}} & -\frac{1}{\sqrt{6}}
\end{bmatrix}
\label{eq:svd_lossyDivider}
\end{equation}

\noindent where it is easy to verify that $\boldsymbol{U}$ and $\boldsymbol{V}$ are real unitary matrices since $\boldsymbol{S}$ has only real elements. Next, we construct an absorption matrix $\boldsymbol{A}$ that is factorized by the same unitary matrices as in $\boldsymbol{S}$. First, we find by definition that $\boldsymbol{\Sigma}_{A}=\frac{\sqrt{3}}{2}diag\left\{ 0,1,1\right\}$, leading to
\begin{equation}
\boldsymbol{A}=\boldsymbol{U}\boldsymbol{\Sigma}_{A}\boldsymbol{V}^{\dagger}=\frac{1}{2\sqrt{3}}\begin{bmatrix}-2 & 1 & 1\\
1 & -2 & 1\\
1 & 1 & -2
\end{bmatrix}
\end{equation}

Finally, substituting $\boldsymbol{S}$ and $\boldsymbol{A}$ into Eq.~(\ref{eq:Delta_SVD}) leads to a complete form for the $\boldsymbol{\varLambda}$ matrix. In this manner, by using the transformation rules given in Eqs.~(\ref{eq:a_dagger})-(\ref{eq:f_dagger}), it is possible to evaluate any quantum state transformation introduced by a T-junction power divider.

By comparing the column vectors of $\boldsymbol{U}$ and $\boldsymbol{V}$, we find that they either are equal or differ by a -1 factor. Due to the similarity of the basis expanded by the $\boldsymbol{U}$ and $\boldsymbol{V}$ matrices, the description of the device provided with the SVD in this case must have very close relationship with an eigenvector decomposition. Since all real symmetric matrices are normal matrices, we can illustrate how this is the case by factoring the scattering matrix of the T-junction power divider with a unitary diagonalization, i.e., $\boldsymbol{S}=\boldsymbol{U}_{\lambda}\boldsymbol{\Sigma}_{\lambda}\boldsymbol{U}_{\lambda}^{\dagger}$. For this particular case, the $\boldsymbol{\Sigma}_{\lambda}$ and $\boldsymbol{U}_{\lambda}$ matrices are given by
\begin{equation}
\boldsymbol{U}_{\lambda}=\begin{bmatrix}\frac{1}{\sqrt{3}} & -\frac{1}{\sqrt{2}} & -\frac{1}{\sqrt{6}}\\
\frac{1}{\sqrt{3}} & 0 & \sqrt{\frac{2}{3}}\\
\frac{1}{\sqrt{3}} & \frac{1}{\sqrt{2}} & -\frac{1}{\sqrt{6}}
\end{bmatrix}\qquad\boldsymbol{\Sigma}_{\lambda}=\begin{bmatrix}1 & 0 & 0\\
0 & -\frac{1}{2} & 0\\
0 & 0 & -\frac{1}{2}
\end{bmatrix}
\label{eq:UDiagon_lossyDivider}
\end{equation}

As predicted, $\boldsymbol{\Sigma}_{A\lambda}=\boldsymbol{\Sigma}_{A}=\frac{\sqrt{3}}{2}diag\left\{ 0,1,1\right\}$ and the absorption matrix associated with the unitary diagonalization of $\boldsymbol{S}$ can then be defined as
\begin{equation}
\boldsymbol{A}_{\lambda}=\boldsymbol{U}_{\lambda}\boldsymbol{\Sigma}_{A\lambda}\boldsymbol{U}_{\lambda}^{\dagger}=\frac{1}{2\sqrt{3}}\begin{bmatrix}2 & -1 & -1\\
-1 & 2 & -1\\
-1 & -1 & 2
\end{bmatrix}
\end{equation}

Analogously to the SVD case, substitution of $\boldsymbol{S}$ and $\boldsymbol{A}_{\lambda}$ in Eq.~(\ref{eq:Delta_UDiagon}) leads to an alternative form for the $\boldsymbol{\varLambda}$ matrix, that could also be used to evaluate any quantum state transformation.

Comparison of the decompositions in Eq.~(\ref{eq:svd_lossyDivider}) and Eq.~(\ref{eq:UDiagon_lossyDivider}) allows us to review the simple connection between an SVD decomposition and a unitary diagonalization. Specifically, the singular values in the diagonal entries of $\boldsymbol{\Sigma}_{S}$ are the absolute values of the eigenvalues, which are the diagonal entries of $\boldsymbol{\Sigma}_{\lambda}$. It is also evident that $\boldsymbol{U}=\boldsymbol{U}_{\lambda}\boldsymbol{\Sigma}_{\vartheta}$, with $\boldsymbol{\Sigma}_{\vartheta}=diag\left\{ 1,-1,-1\right\}$ , revealing how the SVD decomposition could be readily obtained from the unitary diagonalization. Finally, it is interesting to note that in this example $\boldsymbol{A}=-\boldsymbol{A}_{\lambda}$, making it clear that the choice in the factorization of the $\mathbf{S}$ matrix only results in a trivial change in the basis describing the internal modes of the device.

\subsubsection{Wilkinson power divider}

\begin{figure}[!t]
\includegraphics[width=3.35in]{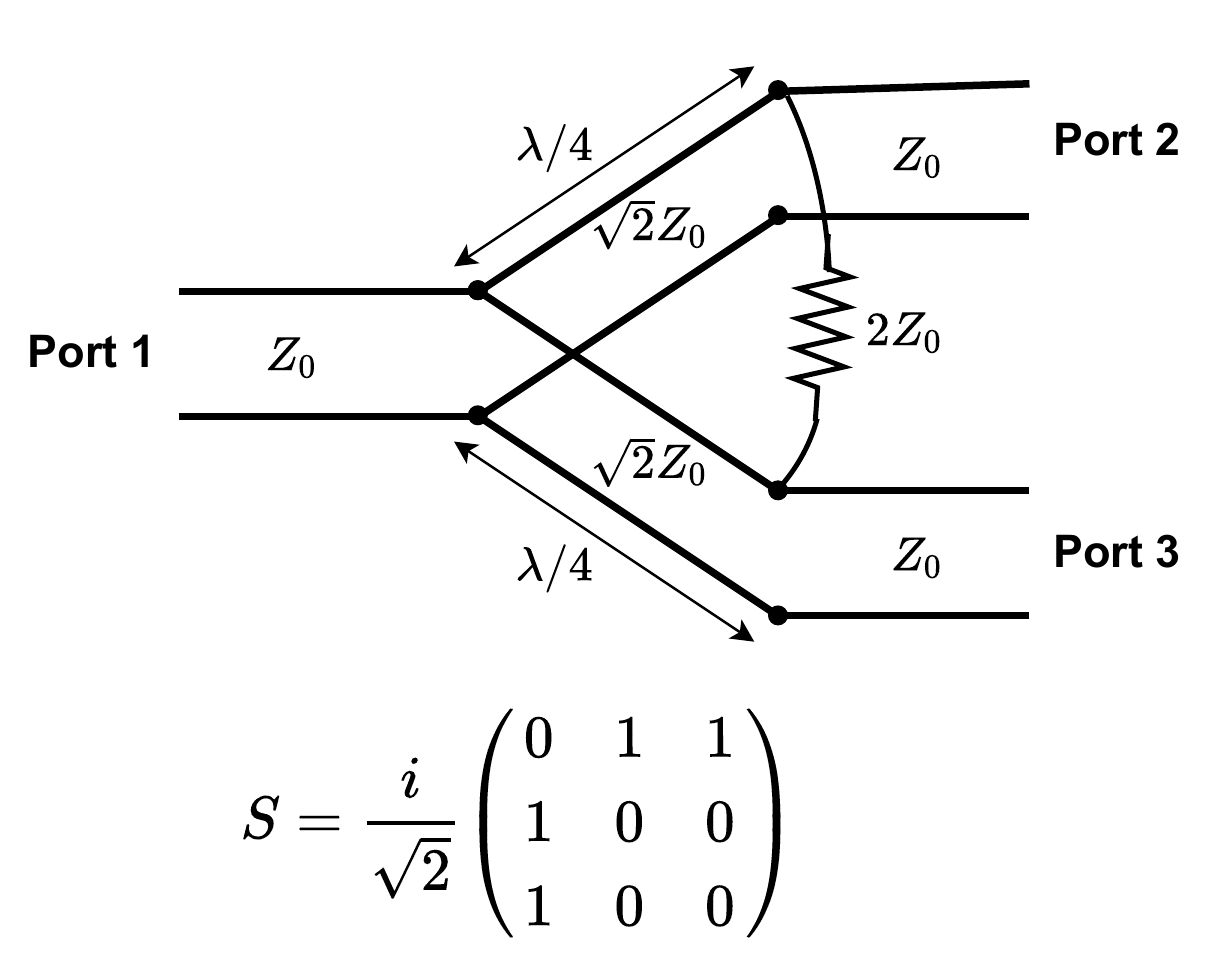}
\caption{Transmission line circuit and scattering matrix of a Wilkinson power divider with an equal power splitting ratio.}
\label{fig:Wilkinson_divider}
\end{figure}

The next reciprocal device ($S_{ij}=S_{ji}$) we analyze to compute its $\boldsymbol{\varLambda}$ matrix is the Wilkinson power divider \cite{pozar2011microwave}. A schematic representation of the device and its scattering matrix are reported in Fig.~\ref{fig:Wilkinson_divider}. It can be drawn from its scattering matrix that, just as the lossy T-junction power divider, the Wilkinson power divider is a three-port reciprocal network with all its ports matched ($S_{ii}=0$). The main difference is that the Wilkinson divider guarantees isolation between its output ports ($S_{23}=S_{32}=0$), as well as power splitting and combining capabilities with unit efficiency ($S_{12}=S_{13}=1/\sqrt{2}$). However, these properties come at the cost of not having the same response when the device is operated in different directions. 

%%%%%%%%%%%%%%%%%%%%%%%%%%%%%%%%%% Figure 3 %%%%%%%%%%%%%%%%%%%%%%%%%%%%%%%%%%%%%%%%%%%%%%%%%%%%%%%%%%%%

%%%%%%%%%%%%%%%%%%%%%%%%%%%%%%%%%%%%%%%%%%%%%%%%%%%%%%%%%%%%%%%%%%%%%%%%%%%%%%%%%%%%%%%%%%%%%%%%%%%%%%%%%

For the scattering matrix in Fig.~\ref{fig:Wilkinson_divider}, a singular value decomposition, i.e., $\boldsymbol{S}=\boldsymbol{U}\boldsymbol{\Sigma}_{S}\boldsymbol{V}^{\dagger}$, is given by the following matrices
\begin{equation}
\boldsymbol{U}=\frac{i}{\sqrt{2}}\begin{bmatrix}\sqrt{2} & 0 & 0\\
0 & 1 & -i\\
0 & 1 & i
\end{bmatrix}\qquad\boldsymbol{\Sigma}_{S}=\begin{bmatrix}1 & 0 & 0\\
0 & 1 & 0\\
0 & 0 & 0
\end{bmatrix}\nonumber
\end{equation}
\begin{equation}
\boldsymbol{V}=\frac{1}{\sqrt{2}}\begin{bmatrix}0 & \sqrt{2} & 0\\
1 & 0 & -1\\
1 & 0 & 1
\end{bmatrix}
\label{eq:svd_WilkinsonDivider}
\end{equation}

In this case, the SVD decomposition provides a very intuitive description of how the device operates. Specifically, comparing the first columns of $\boldsymbol{U}$ and $\boldsymbol{V}$ we find that when the device is symmetrically excited in the second and third ports, all the power is combined into the first port. On the other hand, looking at the second column of $\boldsymbol{U}$ and $\boldsymbol{V}$ shows the if the device is excited from port 1, the power is equally divided into ports two and three. Finally, the analysis of the third column shows that the out-of-phase excitations of ports 2 and 3 leads to perfect absorption of the input power, i.e., coherent perfect absorption (CPA). 

Next, we compute a form of the absorption matrix $\boldsymbol{A}$ that shares a decomposition with the same unitary matrices as $\boldsymbol{S}$. The matrix $\boldsymbol{A}$ is then given by
\begin{equation}
\boldsymbol{A}=\boldsymbol{U}\boldsymbol{\Sigma}_{A}\boldsymbol{V}^{\dagger}=\frac{1}{2}\begin{bmatrix}0 & 0 & 0\\
0 & 1 & -1\\
0 & -1 & 1
\end{bmatrix}
\label{eq:A_SVD_Willkinson}
\end{equation}

\noindent where $\boldsymbol{\Sigma}_{A}=diag\left\{ 0,0,1\right\}$. With this information, the input-output relations can be identified by constructing the $\boldsymbol{\varLambda}$ matrix of the device by substitution of $\boldsymbol{S}$ and $\boldsymbol{A}$ into Eq.~(\ref{eq:Delta_SVD}).

It can be readily verified that the $\boldsymbol{S}$ matrix for the Wilkinson power divider is normal. Therefore, a Wilkinson power divider can also be analyzed with a unitary diagonalization. Specifically, the scattering matrix is factorized as $\boldsymbol{S}=\boldsymbol{U}_{\lambda}\boldsymbol{\Sigma}_{\lambda}\boldsymbol{U}_{\lambda}^{\dagger}$, with the following matrices
\begin{equation}
\boldsymbol{U}_{\lambda}=\begin{bmatrix}-\frac{1}{\sqrt{2}} & \frac{1}{\sqrt{2}} & 0\\
\frac{1}{2} & \frac{1}{2} & -\frac{1}{\sqrt{2}}\\
\frac{1}{2} & \frac{1}{2} & \frac{1}{\sqrt{2}}
\end{bmatrix}\qquad\boldsymbol{\Sigma}_{\lambda}=\begin{bmatrix}-i & 0 & 0\\
0 & i & 0\\
0 & 0 & 0
\end{bmatrix}
\label{eq:UDiagon_Wilkinson}
\end{equation}

As expected, $\boldsymbol{\Sigma}_{A\lambda}=\boldsymbol{\Sigma}_{A}=diag\left\{ 0,0,1\right\}$, and the absorption matrix in this decomposition is given by 
\begin{equation}
\boldsymbol{A}_{\lambda}=\boldsymbol{U}_{\lambda}\boldsymbol{\Sigma}_{A\lambda}\boldsymbol{U}_{\lambda}^{\dagger}=\frac{1}{2}\begin{bmatrix}0 & 0 & 0\\
0 & 1 & -1\\
0 & -1 & 1
\end{bmatrix}
\end{equation}

In this case, the SVD and the unitary diagonalization provide very different perspectives of the same device. While the SVD describes the mode of opertation of the Wilkinson power divider as a transmission device, the unitary diagonalization emphasize those input combinations that are simply scaled by the interaction with the device. At the same time, this example allow us to illustrate that the SVD and unitary diagonalizations are not unique decompositions. In fact, we recall that we can obtain an alternative SVD decomposition from a unitary diagonalization by extracting the phases of $\boldsymbol{\Sigma}_{\lambda}=\boldsymbol{\Sigma}_{\left|\lambda\right|}\boldsymbol{\Sigma}_{\theta}$. However, doing this excercise with the unitary diagonalization above does not lead to the same SVD decomposition reported in Eq.~(\ref{eq:A_SVD_Willkinson}). 

\subsection{Nonreciprocal devices}

\subsubsection{Lossy circulator}

Next, we extend our analysis to nonreciprocal components (having nonsymmetric scattering matrices) to show that the proposed method to compute the $\boldsymbol{\varLambda}$ matrix can be applied to any linear device. As a first example, we consider a lossy circulator. The circulator is a nonreciprocal three-port device with a simple operating principle: the input power from one port flows to the adjacent one, depending on the direction of rotation determined in its design \cite{pozar2011microwave}. Following this functioning principle, we have modeled a lossy circulator, whose schematic representation and scattering matrix are depicted in Fig.4. The transmission coefficients are $a=\left|a\right|e^{i\vartheta_{a}}$, $ b=\left|b\right|e^{i\vartheta_{b}}$, $c=\left|c\right|e^{i\vartheta_{c}}$ with $\left|a\right|,\left|b\right|,\left|c\right|\leq1$.

%%%%%%%%%%%%%%%%%%%%%%%%%%%%%%%%%%%% Figure 4 %%%%%%%%%%%%%%%%%%%%%%%%%%%%%%%%%%%%%%%%%%%%%%%%%%%%%%%%%%

\begin{figure}[!t]
\includegraphics[width=3.35in]{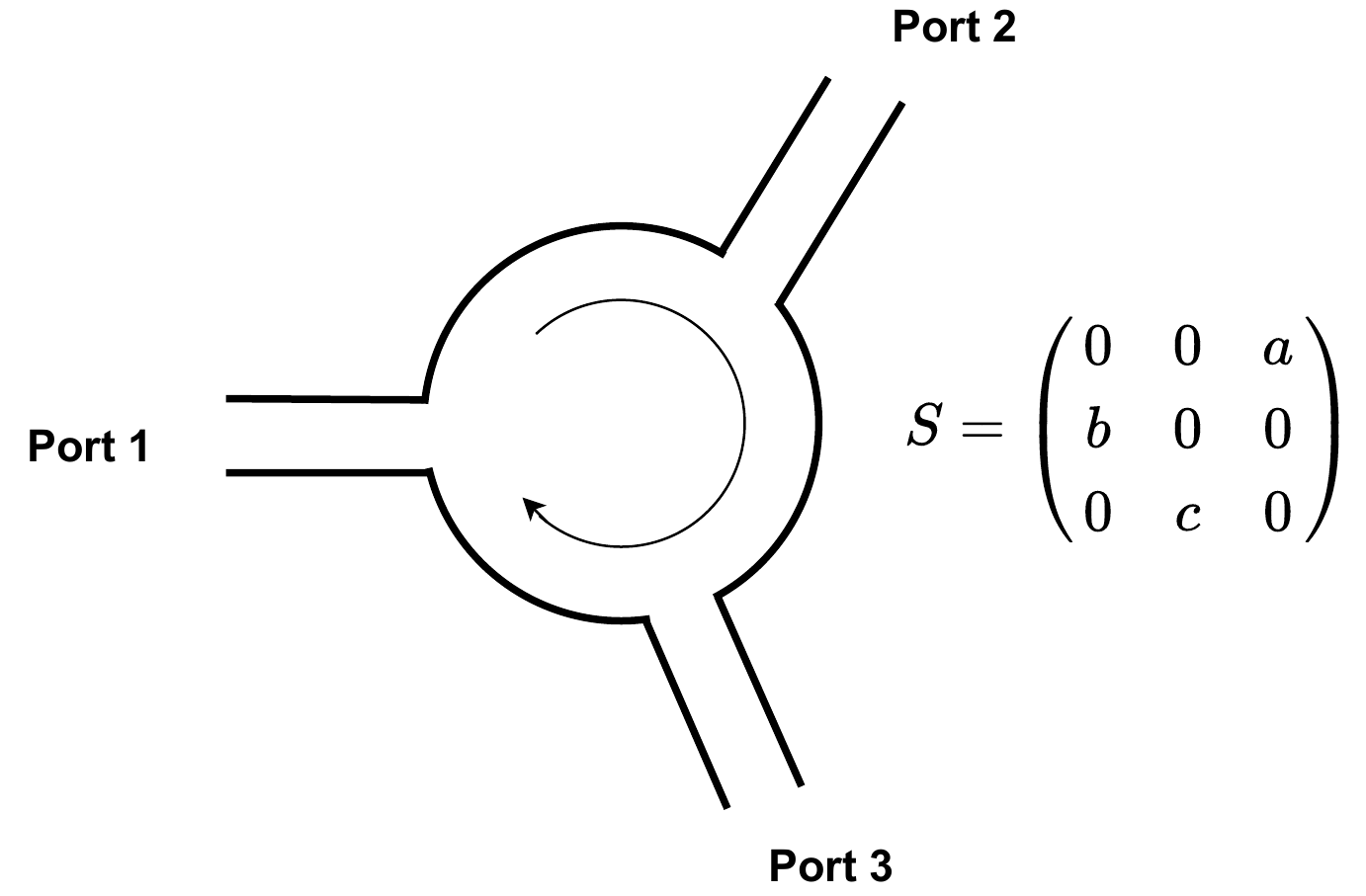}
\caption{Schematic representation and scattering matrix of a circulator.}
\label{fig:Lossy_circulator}
\end{figure}

%%%%%%%%%%%%%%%%%%%%%%%%%%%%%%%%%%%%%%%%%%%%%%%%%%%%%%%%%%%%%%%%%%%%%%%%%%%%%%%%%%%%%%%%%%%%%%%%%%%%%%%%

To obtain the $\boldsymbol{\varLambda}$ matrix of the lossy circulator we must start by performing a singular value decomposition of the scattering matrix in Fig.~\ref{fig:Lossy_circulator}, i.e., $\boldsymbol{S}=\boldsymbol{U}\boldsymbol{\Sigma}_{S}\boldsymbol{V}^{\dagger}$. This can be accomplished through the matrices
\begin{equation}
\boldsymbol{U}=\begin{bmatrix}e^{i\vartheta_{a}} & 0 & 0\\
0 & e^{i\vartheta_{b}} & 0\\
0 & 0 & e^{i\vartheta_{c}}
\end{bmatrix}\qquad\boldsymbol{\Sigma}_{S}=\begin{bmatrix}\left|a\right| & 0 & 0\\
0 & \left|b\right| & 0\\
0 & 0 & \left|c\right|
\end{bmatrix}\nonumber
\end{equation}

\begin{equation}
\boldsymbol{V}=\begin{bmatrix}0 & 1 & 0\\
0 & 0 & 1\\
1 & 0 & 0
\end{bmatrix}
\label{eq:svd_LossyCirculator}
\end{equation}

Again, being mostly a transmission device, the SVD decomposition provides an intuitive picture of the operating principle. Comparing the columns vectors of the $\boldsymbol{U}$ and $\boldsymbol{V}$ matrices it is easy to identify that one port is individually connected to the next port. The diagonal elements of $\boldsymbol{\Sigma}_{S}$ coincide with the absolute values of the transmission coefficients $a$, $b$, and $c$, while $\boldsymbol{U}$ is also a diagonal matrix whose diagonals entries are the phases of those transmission coefficients. 

To complete the generalized input-output relations, we compute a form of the absorption matrix $\boldsymbol{A}$ with a similar decomposition as $\boldsymbol{S}$:
\begin{equation}
\boldsymbol{A}=\boldsymbol{U}\boldsymbol{\Sigma}_{A}\boldsymbol{V}^{\dagger}\nonumber
\end{equation}
\begin{equation}
=\begin{bmatrix}0 & 0 & e^{i\vartheta_{a}}\sqrt{1-\left|a\right|^{2}}\\
e^{i\vartheta_{b}}\sqrt{1-\left|b\right|^{2}} & 0 & 0\\
0 & e^{i\vartheta_{c}}\sqrt{1-\left|c\right|^{2}} & 0
\end{bmatrix}
\label{eq:A_SVD_LossyCirculator}
\end{equation}

\noindent where $\boldsymbol{\Sigma}_{A}=diag\left\{ \sqrt{1-\left|a\right|^{2}},\sqrt{1-\left|b\right|^{2}},\sqrt{1-\left|c\right|^{2}}\right\}$. Finally, we must substitute $\boldsymbol{S}$ and $\boldsymbol{A}$ into Eq.~(\ref{eq:Delta_SVD}) to obtain the $\boldsymbol{\varLambda}$ matrix that fully described the input-output relations, and allows for the analysis of any quantum state transformation that might take place in a lossy circulator.

In general, a lossy circulator does not have a normal matrix, and it cannot be described through a unitary diagonalization. Since circulators are ubiquitous in many optical setups, it is a good example to show how the generality provided by the SVD might be required in many practical scenarios. However, a unitary diagonalization of the a lossy circulator is still possible in the particular case in which the three nonzero transmission coefficients have the same absolute value, i.e., $\left|a\right|=\left|b\right|=\left|c\right|$. With this restriction the scattering matrix of the circulator becomes normal. For the sake of simplicity, we consider a circulator with the same transmission properties (both magnitude and phase) for all its ports, i.e., $a=b=c$. A unitary diagonalization of the $\boldsymbol{S}$ matrix of the uniform lossy circulator is given by the matrices
\begin{equation}
\boldsymbol{U}_{\lambda}=\frac{1}{\sqrt{3}}\begin{bmatrix}1 & e^{-i\frac{2\pi}{3}} & e^{i\frac{2\pi}{3}}\\
1 & e^{i\frac{2\pi}{3}} & e^{-i\frac{2\pi}{3}}\\
1 & 1 & 1
\end{bmatrix}\qquad\boldsymbol{\Sigma}_{\lambda}=a\begin{bmatrix}1 & 0 & 0\\
0 & e^{i\frac{2\pi}{3}} & 0\\
0 & 0 & e^{-i\frac{2\pi}{3}}
\end{bmatrix}
\label{eq:UDiagon_LossyCirculator}
\end{equation}

The unitary diagonalization of the circulator provides and interesting perspective on its mode of operation. Beyond the port-to-port transmission picture, the action of a circulator can be visualized as the circular shift of the elements of a vector by one position. Therefore, eigenvectors must correspond to vectors with the property that a circular shift is equivalent to the multiplication by a scalar factor. It is found in (\ref{eq:UDiagon_LossyCirculator}) that the elements of the eigenvectors correspond to points on a circumference of radius $\sqrt{3}$ in the complex plane, angularly separated by a phase factor $e^{i\frac{2\pi}{3}}$. In this manner, the circular shift of the vector corresponds to the multiplication by scalar phase factors $e^{\pm i\frac{2\pi}{3}}$, playing the role of eigenvalues. On the the other hand, the absorption matrix $\boldsymbol{A}_{\lambda}$ takes the simpler form
\begin{equation}
\boldsymbol{A}_{\lambda}=\boldsymbol{U}_{\lambda}\boldsymbol{\Sigma}_{A\lambda}\boldsymbol{U}_{\lambda}^{\dagger}=\sqrt{1-\left|a\right|^{2}}\begin{bmatrix}1 & 0 & 0\\
0 & 1 & 0\\
0 & 0 & 1
\end{bmatrix}
\end{equation}

\noindent with $\boldsymbol{\Sigma}_{A\lambda}=\boldsymbol{A}_{\lambda}$. This simple diagonal form illustrates that light is separatedly dissipated into different modes, with no coherent effects. Finally, substitution of $\boldsymbol{S}$ and $\boldsymbol{A}_{\lambda}$ in Eq.~(\ref{eq:Delta_UDiagon}) gives us an alternative form for the $\boldsymbol{\varLambda}$ matrix of a uniform lossy circulator.

The study of the lossy circulator remarks the convenience of an SVD decomposition of $\boldsymbol{S}$ to compute the $\boldsymbol{\varLambda}$ matrix. It emphasizes how it describes the behavior of transmission devices, and it shows that it can be applied to general nonreciprocal devices.

\subsubsection{Asymmetric transmission devices}

\begin{figure}[!t]
\includegraphics[width=3.35in]{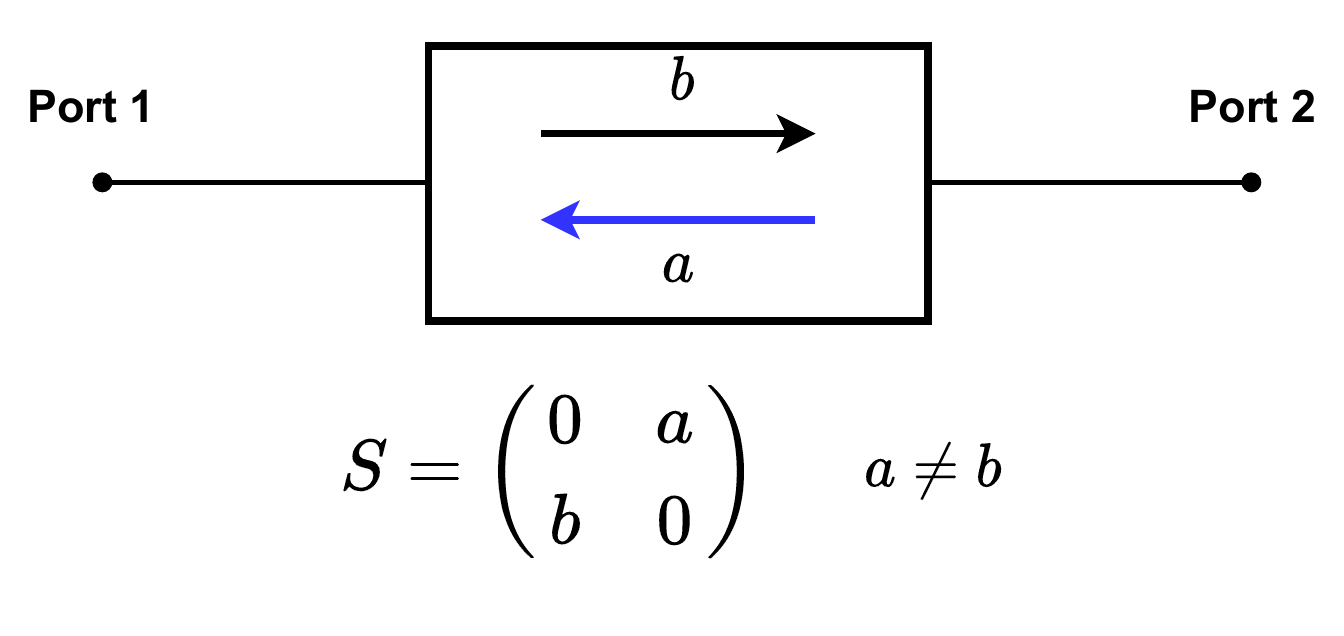}
\caption{Schematic representation and scattering matrix of an asymmetric transmission device.}
\label{fig:AsymmetricTransmissionDevices}
\end{figure}

The final example we study comprises all lossy devices whose functionalities can be described through the schematic representation and scattering matrix depicted in Fig.~\ref{fig:AsymmetricTransmissionDevices}. Such devices consist of matched two-port devices with asymmetric transmission coefficients $a=\left|a\right|e^{i\vartheta_{a}}$ and $b=\left|b\right|e^{i\vartheta_{b}}$. By definition, these are nonreciprocal devices. In the limiting case where $a=1$ and $b=0$ (or $a=0$ and $b=1$), the scattering matrix in Fig.~\ref{fig:AsymmetricTransmissionDevices} represents an ideal isolator: a device perfectly transparent in one direction, while being a perfect absorber from the reverse direction \cite{pozar2011microwave}. On the other hand, if $b=-a$ with $a\neq1$, the resulting scattering matrix describes a lossy gyrator, i.e., the lossy version of the ideal gyrator which has a $180^{\circ}$ differential phase shift \cite{pozar2011microwave}.

%%%%%%%%%%%%%%%%%%%%%%%%%%%%%%%%%%%%%%%%% Figure 5 %%%%%%%%%%%%%%%%%%%%%%%%%%%%%%%%%%%%%%%%%%%%%%%%%%%%%%
%%%%%%%%%%%%%%%%%%%%%%%%%%%%%%%%%%%%%%%%%%%%%%%%%%%%%%%%%%%%%%%%%%%%%%%%%%%%%%%%%%%%%%%%%%%%%%%%%%%%%%%%%

Using the SVD decomposition, the scattering matrix is factorized as $\boldsymbol{S}=\boldsymbol{U}\boldsymbol{\Sigma}_{S}\boldsymbol{V}^{\dagger}$, with
\begin{equation}
\boldsymbol{U}=\begin{bmatrix}e^{i\vartheta_{a}} & 0\\
0 & e^{i\vartheta_{b}}
\end{bmatrix}\qquad\boldsymbol{\Sigma}_{S}=\begin{bmatrix}\left|a\right| & 0\\
0 & \left|b\right|
\end{bmatrix}\qquad\boldsymbol{V}=\begin{bmatrix}0 & 1\\
1 & 0
\end{bmatrix}
\label{eq:svd_AsymmTransmDevices}
\end{equation}

Similar to the circulator, the asymmetric transmission device basically operates in two different transmission modes, and its operating principle is intuitively described by inspecting the column vectors of the $\boldsymbol{U}$ and $\boldsymbol{V}$ matrices in (\ref{eq:svd_AsymmTransmDevices}). Next, we find a form of the absorption matrix $\boldsymbol{A}$, factorized by the same unitary matrices $\boldsymbol{U}$ and $\boldsymbol{V}$:
\begin{equation}
\boldsymbol{A}=\boldsymbol{U}\boldsymbol{\Sigma}_{A}\boldsymbol{V}^{\dagger}=\begin{bmatrix}0 & e^{i\vartheta_{a}}\sqrt{1-\left|a\right|^{2}}\\
e^{i\vartheta_{b}}\sqrt{1-\left|b\right|^{2}} & 0
\end{bmatrix}
\label{eq:A_SVD_AsymmTransmDevices}
\end{equation}

\noindent where $\boldsymbol{\Sigma}_{A}=diag\left\{ \sqrt{1-\left|a\right|^{2}},\sqrt{1-\left|b\right|^{2}}\right\}$. Then, substituting $\boldsymbol{S}$ and $\boldsymbol{A}$ into Eq.~(\ref{eq:Delta_SVD}) leads us to the $\boldsymbol{\varLambda}$ matrix providing the input-output relations to describe quantum states transformations produced by two-port asymmetric transmission devices.

The alternative approach to compute the $\boldsymbol{\varLambda}$ matrix via the unitary diagonalization of $\boldsymbol{S}$ is only possible if $\left|a\right|=\left|b\right|$. Since the $\left|a\right|=\left|b\right|$ condition implies a symmetric magnitude transition, it suggests that an SVD will be preferred for most asymmetric transmission devices. However, a unitary diagonalization of the $\boldsymbol{S}$ matrix is still possible for the particular case of a lossy gyrator, i.e., $\boldsymbol{S}=\boldsymbol{U}_{\lambda}\boldsymbol{\Sigma}_{\lambda}\boldsymbol{U}_{\lambda}^{\dagger}$, as $b=-a$. In fact, a unitary diagonalization of the $\boldsymbol{S}$ matrix of a lossy gyrator is given by
\begin{equation}
\boldsymbol{U}_{\lambda}=\frac{1}{\sqrt{2}}\begin{bmatrix}-i & i\\
1 & 1
\end{bmatrix}\qquad\boldsymbol{\Sigma}_{\lambda}=a\begin{bmatrix}i & 0\\
0 & -i
\end{bmatrix}
\label{eq:UDiagon_LossyGyrator}
\end{equation}

\noindent while the absorption matrix $\boldsymbol{A}_{\lambda}$ can be written as follows:
\begin{equation}
\boldsymbol{A}_{\lambda}=\boldsymbol{U}_{\lambda}\boldsymbol{\Sigma}_{A\lambda}\boldsymbol{U}_{\lambda}^{\dagger}=\sqrt{1-\left|a\right|^{2}}\begin{bmatrix}1 & 0\\
0 & 1
\end{bmatrix}
\end{equation}

\noindent with $\boldsymbol{\Sigma}_{A\lambda}=\boldsymbol{A}_{\lambda}$. Finally, we must substitute $\boldsymbol{S}$ and $\boldsymbol{A}$ into Eq.~(\ref{eq:Delta_UDiagon}) to obtain the associated $\boldsymbol{\varLambda}$ matrix.

The analyzed examples highlight, once again, how applying an SVD of the scattering matrix we can compute the $\boldsymbol{\varLambda}$ matrix of any lossy device, including nonreciprocal devices with nilpotent matrices. Another major advantage of the proposed method, as mentioned earlier, is that SVD routines are available in commonly used mathematical software packages \cite{Matlab, Mathematica}.

\section{Conclusions}

A general procedure to compute the input-output relations characterizing quantum interference in lossy linear devices has been presented. The proposed method is based on a singular value decomposition (SVD) of the scattering matrix $\boldsymbol{S}$, and it can be applied to any type of linear device. In fact, it can be evaluated with popular mathematical software packages, so that evaluating input-output relations becomes a very simple task. An alternative decomposition based on a unitary diagonalization of $\boldsymbol{S}$ has also been proposed. This procedure allows for a description in terms of the eigenvectors of the scattering matrix, but its applicability is restricted to devices with a normal scattering matrix. Both approaches, the SVD decomposition, and the unitary diagonalization, provide different but intuitive and even complementary perspectives on the principle of operation of lossy linear devices. The efficacy of the procedure was demonstrated by computing, with both approaches where possible, input-output relations of popular lossy linear devices. Our examples have included both reciprocal and nonreciprocal devices, such as T-junction and Wilkinson power dividers, circulators and asymmetric transmission devices. We expect that our results will motivate the analysis of quantum interference in complex lossy $N$-port networks. It will allow for carefully evaluating the impact of loss in linear optics quantum technologies, as well as for discovering novel quantum interference phenomena in advanced devices.

\begin{acknowledgments}
I.L. acknowledges support from Ram\'on y Cajal fellowship RYC2018-024123-I and project RTI2018-093714-301J-I00 sponsored by MCIU/AEI/FEDER/UE, and ERC Starting Grant 948504. The authors thank interesting discussions with A. Moreno-Pe\~narrubia.
\end{acknowledgments}

\bibliography{library}% Produces the bibliography via BibTeX.

\end{document}